\documentclass[12pt,preprint,flushrt]{aastex}
\begin{document}

\title{X-ray Bursts from the Accreting Millisecond Pulsar XTE J1814-338}
\author{Tod E. Strohmayer$^1$, Craig B. Markwardt$^2$, Jean H. Swank$^1$, and 
Jean in 't Zand$^3$}
\altaffiltext{1}{Laboratory for High Energy Astrophysics, NASA's Goddard Space 
Flight Center, Greenbelt, MD 20771 email: stroh, swank@milkyway.gsfc.nasa.gov}
\altaffiltext{2}{LHEA/University of Maryland, College Park, MD 20772 email: 
craigm@milkyway.gsfc.nasa.gov}
\altaffiltext{3}{SRON National Institute for Space Research, Sorbonnelaan 2,
NL - 3584 CA Utrecht, the Netherlands, email: jeanz@sron.nl}

\begin{abstract}
 
Since the discovery of the accreting millisecond pulsar XTE J1814-338
a total of 27 thermonuclear bursts have been observed from the source
with the Proportional Counter Array (PCA) onboard the Rossi X-ray
Timing Explorer (RXTE).  Spectroscopy of the bursts, as well as the
presence of continuous burst oscillations, suggests that all but one
of the bursts are sub-Eddington. The remaining burst has the largest
peak bolometric flux of $2.64 \times10^{-8}$~erg~s$^{-1}$cm$^{-2}$, as
well as a gap in the burst oscillations, similar to that seen in
Eddington limited bursts from other sources.  Assuming this burst was
Eddington limited we obtain a source distance of $\approx 8$ kpc. All
the bursts show coherent oscillations at the 314.4 Hz spin
frequency. The burst oscillations are strongly frequency and phase
locked to the persistent pulsations. Only two bursts show evidence for
frequency drift in the first few seconds following burst onset. In
both cases the initial drift corresponds to a spin down of a few
tenths of a Hz. The large oscillation amplitude during the bursts
confirms that the burst flux is modulated at the spin frequency.  We
detect, for the first time, a significant first harmonic component in
burst oscillations. The ratio of countrate in the first harmonic to
that in the fundamental can be $ > 0.25$ and is, on average, less than
that of the persistent pulsations.  If the pulsations result from a
single bright region on the surface, the harmonic strength suggests
the burst emission is beamed, perhaps due to a stronger magnetic field
than in non-pulsing LMXBs.  Alternatively, the harmonic content could
result from a geometry with two bright regions.
 
\end{abstract}

\keywords{binaries: general -- stars: individual (XTE J1814-338) -- stars: 
neutron -- stars: rotation -- X-rays: bursts - X-rays: stars}

\vfill\eject

\section{Introduction}

Recently, Markwardt \& Swank (2003) reported the discovery of the
fifth accreting millisecond pulsar, XTE J1814-338 (hereafter
J1814). This pulsar has a 314.36 Hz spin frequency, resides in a
binary with an orbital operiod of 4.275 hours, and has a minimum
companion mass of $\approx 0.15 M_{\odot}$ (Markwardt et al. 2003).
This is the widest, and most massive binary of the 5 known accreting
millisecond pulsars. The system parameters of J1814 closely match
those of the ``classical'', non-pulsing low mass X-ray binaries
(LMXBs), for example the 3.8 hour binary 4U 1636-53. These systems are
not generally observed as persistent pulsars, but the neutron star
spins are known from the detection of oscillations during
thermonuclear bursts (see Strohmayer \& Bildsten 2003 for a review of
burst oscillations). Evidence is emerging that the accreting
millisecond pulsars must have stronger large scale magnetic fields
than the non-pulsing LMXBs (Galloway et al. 2002; Cumming, Zweibel \&
Bildsten 2001).  The fundamental quantity controlling the field
strength may be the long term accretion rate, if the accretion can
``bury'' the field.  These pulsars are all transients and/or
subluminous compared to other LMXBs, suggesting a lower long term
accretion rate.  Magnetic fields also play a key role in the frequency
stability of burst oscillations (Cumming, Zweibel \& Bildsten
2001). Detailed studies of J1814 and its bursts may thus provide
insight into the magnetic fields of LMXBs, and why some do not show
persistent pulsations.

The recent discoveries of coherent oscillations during a superburst
from 4U 1636-53 (Strohmayer \& Markwardt 2002), and burst oscillations
from the millisecond pulsar SAX J1808.4-3658 (hereafter J1808,
Chakrabarty et al. 2003), have conclusively established that the
frequency of burst oscillations are directly related to the spin of
the neutron star. Since the discovery of J1814 in June, 2003 extensive
observing with RXTE has been undertaken.  At the time of this writing
a total of 27 thermonuclear bursts have been observed.  Pulse trains
at the 314.36 Hz spin frequency have been detected in all of these
bursts, making this source the second for which burst oscillations
have been observed at the known spin frequency of the neutron star.
In this Letter we report on a study of these first bursts observed
from J1814. As we show below, the burst oscillations are phase locked
to the persistent pulsations. We find good evidence for frequency
drifting near the onset of only two of the bursts. Most interestingly,
we detect the first significant harmonic structure in burst
oscillation waveforms. We use spectroscopy of the bursts to obtain a
distance constraint. We conclude with some discussion of the
implications of our findings.

\section{Burst Time Profiles and Spectroscopy}

Some basic properties of a representative sample of the bursts are
given in Table 1. Almost all the bursts last typically 2 minutes and
are characterized by a rise time (as defined by the time it takes the
photon count rate to grow from 10\% to 100\% of the peak value) of 5
to 10 seconds, and $1/e$ decay times, as measured from the peak
onward, of 20 to 35 seconds. Two different temporal components are
present: a bright and fast component with peak rates above
600~c~s$^{-1}${\rm PCU}$^{-1}$ and durations of at most 30 seconds,
and a slow component that is responsible for the overall duration.
The weak bursts appear to lack the fast component. The timescale of
the longer component suggests that the neutron star is accreting
hydrogen-rich material at a rate where stable hydrogen burning is
either not occurring (implying $\dot m < 9\times10^2$
g~cm$^{-2}$s$^{-1}$) or the burning rate is too slow to catch up with
the supply of fresh fuel (implying $\dot m > 2\times10^3$
g~cm$^{-2}$s$^{-1}$; Fujimoto, Hanawa \& Miyaji 1981, Fushiki \& Lamb
1987; Cumming \& Bildsten 2000).  The peak count rate, corrected for
pre-burst persistent levels and evaluated with 0.5~second time
resolution, ranges between $\approx$0.4 to 1.4 kcts~s$^{-1}${\rm
PCU}$^{-1}$. Burst 27, the last burst observed, is qualitatively
different from the others in that its rise time is shorter, and it
reached a higher peak flux.  These differences suggest a higher
fraction of helium fuel for this burst.  If stable hydrogen burning
were taking place prior to this burst, then the lower $\dot m$ at this
epoch should have allowed for more of the accumulating hydrogen to
burn, perhaps resulting in a higher helium fraction.

The background-subtracted spectral data could be well fitted with
single temperature black body radiation and a negligible absorbing
column, $n_{\rm H}$. The evolution of the temperature and black body
radius are similar for most bursts, except that the peak temperatures
of the fainter bursts, and burst 27, are $\approx 1.7$ and 3 keV,
while all others are in the range from 2.3 to 2.5 keV.  None of the
bursts convincingly show spectroscopic evidence of photospheric radius
expansion.  Therefore all bursts, with the exception of burst 27,
appear to be sub-Eddington.  This is also consistent with the presence
of burst oscillations throughout the peaks of these bursts (Muno et
al. 2002).  Again, burst 27 is different in that it's burst
oscillations show a gap across the peak (see Figure 1, top
right). Such gaps are seen in many radius expansion bursts with
oscillations (Muno et al. 2002), suggesting that burst 27 reached the
Eddington limit. As noted above, this burst also had the highest peak
color temperature, $> 3$ keV, of any of the bursts. If this burst was
indeed Eddington limited then we get a distance contraint based on the
peak bolometric flux of
2.64$\times10^{-8}$~erg~s$^{-1}$cm$^{-2}$. Assuming an Eddington limit
of 2$\times10^{38}$~erg~s$^{-1}$ (for a canonical neutron star and a
hydrogen dominated photosphere) we get a distance of $\approx 8$
kpc. The uncertainty on this distance is estimated to be roughly 20\%
(Kuulkers et al. 2003).

\subsection{Burst Oscillations}

For all the bursts we have 125 $\mu$-sec time resolution PCA event
mode data. We first barycentered the data using the JPL DE405 solar
system ephemeris and the source position determined from PCA scans
(Markwardt \& Swank 2003). Our timing analysis is performed using the
$Z^2_n$ statistic (Buccheri et al. 1983; Strohmayer \& Markwardt
2002).  We first computed dynamic power spectra for each burst. We
used 4 s intervals to compute $Z_1^2$, and we started a new interval
every 0.25 seconds.  Significant oscillations were detected in each
burst in the vicinity of the 314 Hz pulsar frequency. The pulse trains
during most of the bursts lasted from 30 - 100 seconds. These are
longer than the pulse trains observed from most burst oscillation
sources, and allow the time evolution of the pulsed amplitude to be
explored in some detail. Most bursts show complex, episodic
modulations in the amplitude of $\approx 3 - 5 \%$ (rms) on timescales
ranging from 7 - 15 s.  We will explore this interesting behavior in
more detail in a sequel.  The dynamic spectra indicate that the
frequency during most bursts is stable, showing little evidence for
the drifting seen in other sources (Strohmayer \& Markwardt 1999; Muno
et al. 2002).

We performed a phase coherent timing analysis for each burst. We used
a linear frequency evolution model, $\nu (t) = \nu_0 + \delta\nu t$
(ie. the phase is quadratic in time), to model the phase of the
oscillations. We tracked the phases in successive time bins throughout
each burst and found the values of $\nu_0$ and $\delta\nu$ which gave
the best fit to the data. The best fit is found by minimizing $\chi^2
= \sum_{i=1}^N (\phi_i (\nu_0 , \delta\nu ) - \mu )^2 /
\sigma_{\phi_i}^2$, where $\mu = (1/N)\sum_{i=1}^N \phi_i (\nu_0 ,
\delta\nu )$, $\phi_i$ is the phase deduced for the ith bin using the
model parameters and all X-ray events in the bin, and $N$ is the
number of phase bins.

For each burst the linear frequency model gave an acceptable fit to
the phase timing data, and in no case was there a need for a linear
drift term, $\delta\nu$, larger than the local change in frequency due
to orbital motion. Figure 1 shows dynamic power spectra (top) and
phase timing residuals (bottom) for bursts 11 and 27. These were the
two bursts which showed significant episodes of frequency drift. In
both bursts the frequency evolution shows a spin-down of a few tenths
of Hz near the onset of the burst, followed by a statistically weaker
recovery (spin up).  This behavior appears contrary to that found for
J1808, which showed a rapid, few Hz spin up during the rise of a burst
(Chakrabarty et al. 2003), although the observed spin up in J1808
might conceivably be related to the recovery phase of the evolution
seen in J1814.  Nevertheless, the magnitude of the frequency change
was much larger in J1808 ($\approx 4$ Hz). The observation of spin
downs in J1814 seems atypical of bursters in general, since most
observed drifts are spin ups (see Muno et al. 2002).

We compared the best fitting values for $\nu_0$ for bursts 1 - 12 with
the predicted frequency of the persistent pulsations based on the
orbital solution of Markwardt et al. (2003).  The results for these 12
bursts are shown in Figure 2.  The frequencies of the oscillations
during each burst are consistent with the predicted, orbitally
modulated, pulsar frequency.  Not only are the frequencies consistent,
but the phase of the burst oscillations is strongly locked to that of
the persistent pulsations. In Figure 3 we compare the phase folded
pulse profile from the first 56 ksecs of non-burst data (thick solid
curve) with the co-added pulse profiles from bursts 2 and 3. The burst
data were folded with the same ephemeris as the persistent pulse
data. Both the fundamental and 1st harmonic components of the profile
are aligned. Typical 1$\sigma$ upper limits to any average phase
offsets during bursts for the fundamental and 1st harmonic components
are $\approx 0.7$ and $1.2 \%$, respectively.  These limits are
smaller than the corresponding phase offset for bursts from J1808 of
$\approx 11\%$ (Chakrabarty et al. 2003).

We computed the average amplitudes of the burst oscillations from the
phase folded pulse profiles (see Table 1).  We define the amplitude as
$a = (I_{max} - I_{min}) / (I_{max} + I_{min})$, where $I$ is the
countrate of the folded profile. For these estimates we have not
subtracted off the preburst, persistent emission. Doing so would
increase the amplitudes modestly. The amplitudes are in the range from
14 - 18 \%, except for burst 27, which showed systematicalluy weaker
oscillations. This is larger than the amplitude of the persistent
pulsations, which ranges from about 9 - 12 \% when measured in the
same way after first subtracting off the detector background.  This
confirms that the black body burst flux {\it must} be modulated at the
spin frequency. If it were not the amplitude during the bursts would
be less than 1\%.

The burst oscillations also show a significant harmonic component in
their pulse profiles. We demonstrate this in Figure 4, which shows the
co-added pulse profile from bursts 2 and 3, as well as a model fit
using two Fourier components. The solid curve is the best model, while
the dashed and dotted lines show the fundamental and first harmonic
components.  The harmonic component is strongly required, with an
amplitude of $36 \pm 2.6$ counts s$^{-1}$. Co-adding bursts 2 and 3 in
phase gives $Z_1^2 = 95.2$ at the first harmonic. Since $Z_1^2$ is
distributed as $\chi^2$ with two degrees of freedom, the value of 95.2
is a highly significant detection. The harmonic content varies
significantly from burst to burst. When expressed as the ratio of
pulsed amplitude at the first harmonic to that at the fundamental, it
ranges from essentially 0 -- 0.27. We note that this is always less
than the harmonic content of the persistent pulsations, for which the
same ratio is $\approx 0.33$. In terms of countrate the first harmonic
component for the persistent pulsations around the epoch of bursts 2
and 3 is about 4.5 counts s$^{-1}$, which is much to small to account
for the harmonic content during these bursts.

\section{Discussion and Summary}

J1814 is now the second pulsar to show burst oscillations at the spin
frequency of the neutron star. The large amplitude of the modulations
confirms that the burst flux is spin modulated, as first suggested by
Strohmayer et al. (1996). As in J1808 (Chakrabarty et al. 2003), the
phase of the burst oscillations in J1814 are linked to the phase of
the persistent pulsations. It seems likely that this phasing results
from the magnetic field of the neutron star. In persistent pulsars the
magnetic field must be strong enough to channel the accretion flow and
produce an asymmetric accretion luminosity.  Such a field would then
likely also influence the surface distribution of nuclear fuel,
resulting in a thermonuclear flux distribution during bursts which
tracks the magnetic field geometry. A strong field will also enforce
co-rotation of the surface layers, minimizing shearing and thus
frequency drifting. The fact that the burst oscillation sources which
do not show persistent pulsations also have the frequency drifts with
the longest relaxation time seems consistent with this idea
(Strohmayer \& Markwardt 1999).
 
Chakrabarty et al. (2003) found evidence for frequency drifting at the
onset of a burst from J1808. Although most of the bursts from J1814
that we have examined show no strong evidence of drifting, bursts 11
and 27 are the exceptions.  These bursts show evidence for phase
drifting near burst onset (Figure 2, lower panels). For these bursts
both the phase residuals and the dynamic power spectra suggest that
the oscillation frequency was initially higher than the spin frequency
and then decreased below the spin frequency before recovering again.
For both bursts the evolution takes place in an $\approx 5$ s interval
during the burst rise.

Perhaps the most interesting result is the detection of significant
harmonic content in the burst oscillations.  Two important conclusions
are that the harmonic content must be intrinsic to the modulation of
the thermonuclear flux, and secondly, the harmonic content of the
burst oscillations, expressed as the ratio of power at the first
harmonic to that at the fundamental, is always less than the harmonic
content of the persistent pulsations. As noted by several authors
(Strohmayer et al. 1998; Miller \& Lamb 1998; Weinberg, Miller \& Lamb
2001; Muno et al. 2002), the harmonic content of burst oscillations
can in principle be used to constrain the neutron star radius. This is
because of the relativistic effects caused by the high surface
velocities of these pulsars. The bursts from J1814 with the highest
first harmonic content have a harmonic to fundamental ratio greater
than 0.25. In contrast, upper limits on this ratio in other burst
oscillation sources can be as small as 0.05 (Muno et al.  2002). For a
10 km neutron star, $v_{surf} \approx 0.07 c$ for the spin frequency
of J1814. At this velocity such a large harmonic ratio is extremely
difficult to produce if the emission is isotropic from a single bright
region on the neutron star surface (Weinberg, Miller \& Lamb 2001;
Muno, Ozel \& Chakrabarty 2002). Our results suggest that substantial
beaming of the radiation likely occurs if the modulation is from a
single bright region on the surface. Alternatively, a geometry with
two antipodal bright regions is possible. Detailed modeling of the
energy dependent pulse profiles could provide constraints on the
stellar compactness, radius and perhaps the magnetic field strength
and geometry.


\vfill\eject

\vfill\eject

\section{Figure Captions}

\vskip 10pt

\figcaption[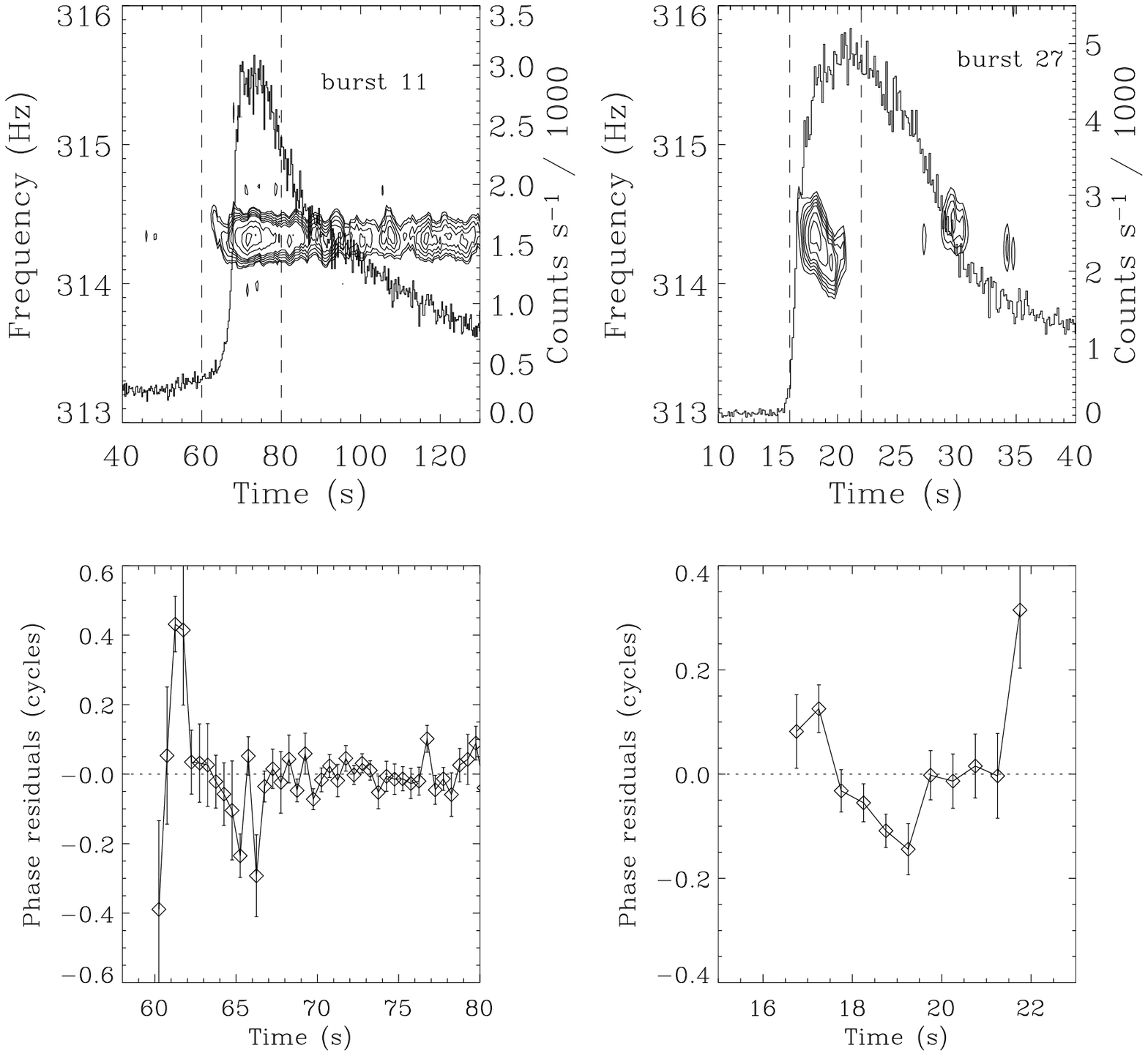]{Summary of timing behavior for two representative
bursts from J1814 (bursts 11 and 27 in Table 1). The top panels show
contour plots of dynamic power spectra computed with the $Z_1^2$
statistic. The time intervals for the spectra were 4 s in length and a
new interval was started every 0.25 s. The bottom panels show the
phase residuals for the best fitting constant frequency model.  Note
the phase modulations near the onset of each burst. In addition to the
gap in the pulse train for burst 27, it also had the weakest
oscillation of any burst and no detected harmonic.
\label{fig1}} 

\vskip 10pt

\figcaption[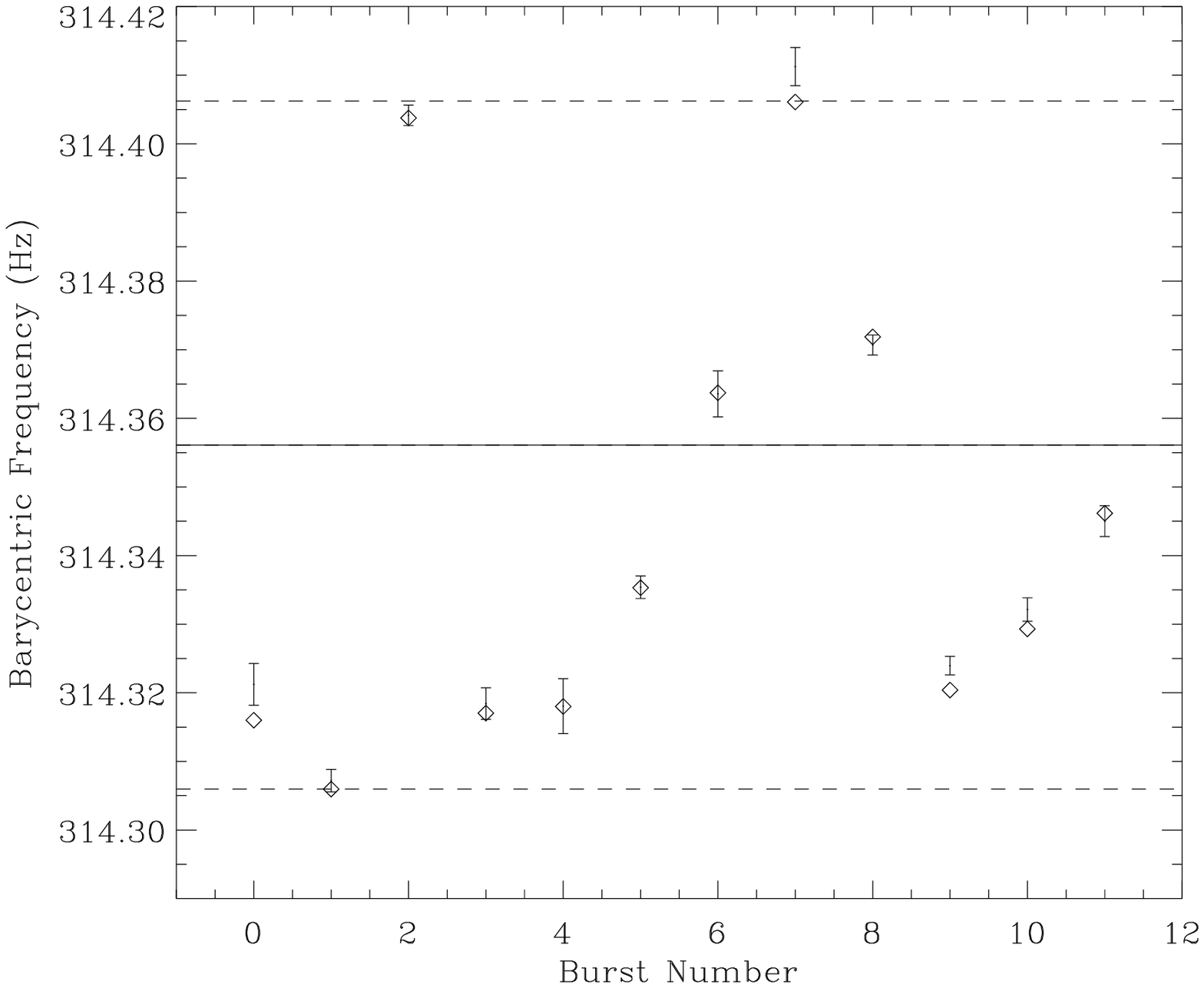]{Comparison of the measured burst oscillation
frequencies (error bar symbols), and the predicted pulsar frequency
based on the orbital Doppler modulation (diamonds). The horizontal and
dashed solid lines mark the pulsar rest frame frequency, and the upper
and lower frequency limits from the binary motion, respectively. The
measured frequencies are consistent with the predicted values.
\label{fig2}} 

\vskip 10pt

\figcaption[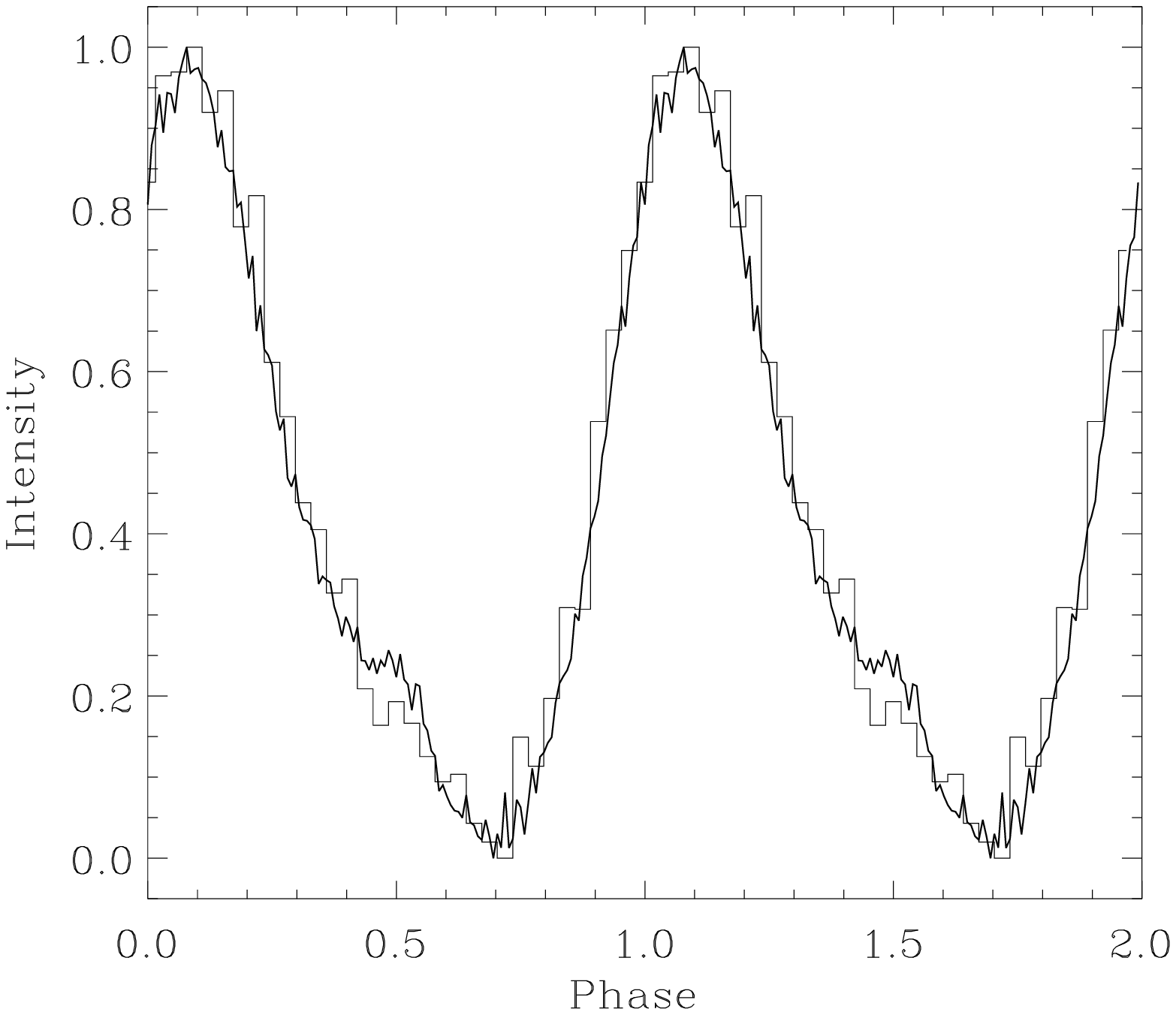]{Comparison of pulse profiles for the persistent
pulsations (solid), and the co-added profiles for bursts 2 and 3
(histogram). The phases of the burst data were computed using the same
orbital ephemeris as for the persistent pulsations. The fact that the
profiles are aligned to a high degree demonstrates that the burst
oscillations are phase locked to the persistent pulsations.
\label{fig3}} 

\vskip 10pt

\figcaption[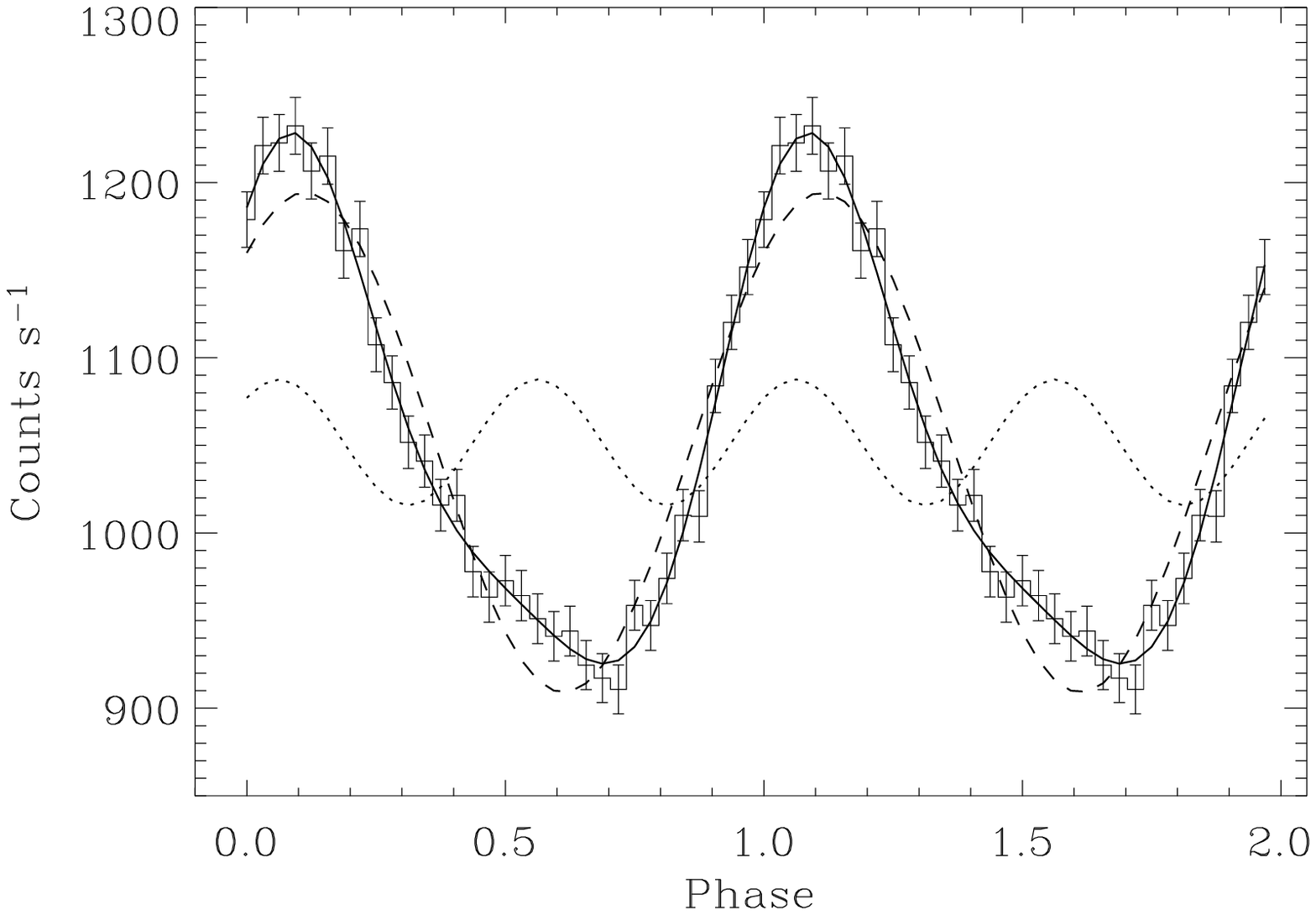]{Co-added pulse profile for the sum of bursts 2 and
3 (histogram with error bars). Also shown are the best fitting, two
Fourier component model (solid), and the individual contributions from
the fundamental (dashed) and first harmonic (dotted) components. The
harmonic component has a strength of $36 \pm 2.6$ counts s$^{-1}$ and
is strongly required by the data. The model gives an excellent fit,
with a minimum $\chi^2 = 19.4$ with 25 degrees of freedom. Two cycles
are shown for clarity.
\label{fig4}} 

\vfill\eject

\newpage

\clearpage

\begin{figure}
\begin{center}
 \includegraphics[width=6in, height=6in]{f1.ps}
\end{center}
Figure 1: Summary of timing behavior for two representative bursts
from J1814 (bursts 11 and 27 in Table 1). The top panels show contour
plots of dynamic power spectra computed with the $Z_1^2$
statistic. The time intervals for the spectra were 4 s in length and a
new interval was started every 0.25 s. The bottom panels show the
phase residuals for the best fitting constant frequency model.  Note
the phase modulations near the onset of each burst. In addition to the
gap in the pulse train for burst 27, it also had the weakest
oscillations of any burst and no detected harmonic.
\end{figure}

\begin{figure}
\begin{center}
 \includegraphics[width=6in,height=5in]{f2.ps}
\end{center}
Figure 2: Comparison of the measured burst oscillation frequencies
(error bar symbols), and the predicted pulsar frequency based on the
orbital Doppler modulation (diamonds). The horizontal and dashed solid
lines mark the pulsar rest frame frequency, and the upper and lower
frequency limits from the binary motion, respectively. The measured
frequencies are consistent with the predicted values.

\end{figure}

\begin{figure}
\begin{center}
 \includegraphics[width=6in, height=5in]{f3.ps}
\end{center}
Figure 3: Comparison of pulse profiles for the persistent pulsations
(solid), and the co-added profiles for bursts 2 and 3 (histogram). The
phases of the burst data were computed using the same orbital
ephemeris as for the persistent pulsations. The fact that the profiles
are aligned to a high degree demonstrates that the burst oscillations
are phase locked to the persistent pulsations.

\end{figure}

\begin{figure}
\begin{center}
 \includegraphics[width=6in, height=5in]{f4.ps}
\end{center}
Figure 4: Co-added pulse profile for the sum of bursts 2 and 3
(histogram with error bars). Also shown are the best fitting, two
Fourier component model (solid), and the individual contributions from
the fundamental (dashed) and first harmonic (dotted) components. The
harmonic component has a strenght of $36 \pm 2.6$ counts s$^{-1}$ and
is strongly required by the data. The model gives an excellent fit,
with a minimum $\chi^2 = 19.4$ with 25 degrees of freedom. Two cycles
are shown for clarity.

\end{figure}

\begin{table}
\caption[]{Spectroscopy and Pulse Timing Results for Bursts from J1814}
\begin{tabular}{cccccccc}
\hline\hline
Burst & Peak        & Peak           & 1/e time & Average & Oscillation & 
Average & Harmonic \cr
index & time$^1$ & flux$^2$ & (sec)$^3$   & radius$^4$ & Frequency$^5$ & 
Amplitude$^6$ & Content$^7$ \cr
\hline
1 &  6.0407 & 0.41$\pm0.05$ & 22.5$\pm0.5$ & 6.75$\pm0.12$ & 314.321 & 0.139 & 
0.040 \\
2 &  7.2687 & 1.25          & 33.3       & 6.71	& 314.307 & 0.148 & 
0.081 \\
3 &  7.8836 & 1.36          & 32.3      & 6.81   & 314.404 & 0.162 & 
0.053 \\
4 &  9.1963 & 1.18          & 32.4       & 6.64  & 314.318 & 0.181 & 
0.036 \\
5 & 10.0993 & 1.44          & 30.9      & 6.25	  & 314.318 & 0.147 & 
0.063 \\
6 & 11.0293 & 0.36          & 25.6       & 6.07   & 314.335 & 0.165 & 
0.047 \\
7 & 12.4665 & 0.87          & 34.3      & 6.33	  & 314.364 & 0.176 & 
0.035 \\
8 & 13.0594 & 0.40          & 28.8       & 6.22    & 314.411 & 0.145 & 
0.078 \\
9 & 13.7342 & 0.95          & 33.2       & 6.36    & 314.371 & 0.156 & 
0.026 \\
10 & 14.0145 & 1.00         & 34.8       & 6.32     & 314.324 & 0.146 & 
0.018 \\
11 & 15.7895 & 0.84         & 35.1       & 6.32     & 314.332 & 0.155 & 
0.015 \\
12 & 16.7475 & 0.85         & 38.4       & 6.20     & 314.345 & 0.177 & 
0.044 \\
27 & 17.7435 & 2.64         & 29.4       & 6.07     & 314.323 & 0.10 & 
 -- \\
\hline\hline
\end{tabular}

$^1$Date in June 2003 (UTC), except for burst 27, which is July 2003 (UTC) \\
$^2$Bolometric, in units of 10$^{-8}$~erg~s$^{-1}$cm$^{-2}$, evaluated with 
4 s resolution \\
$^3$ exponential decay time for the slow component of the burst profile \\
$^4$averaged over 40 sec time interval after the peak of each burst,
in km for a distance of 10~kpc. \\
$^5$Barycentric frequency in Hz. \\
$^6$Average pulsation amplitude, defined as $(I_{max} - I_{min}) / (I_{max} 
+ I_{min})$ \\
$^7$Ratio of $Z^2$ power at the first harmonic to that at the fundamental, if
harmonic detected.
\end{table}

\end{document}